\documentclass[aps,showpacs,nofootinbib]{revtex4-1}
\usepackage{amsmath}
\usepackage{amsfonts}
\usepackage{amssymb,amscd,amsbsy}
\usepackage{epsfig}
\usepackage{color}
\usepackage{mathtools}

\newcommand{\bea}{\begin{eqnarray}}
\newcommand{\eea}{\end{eqnarray}}

\newcommand{\vect}[1]{\mathbf{#1}}

\newcommand{\kt}{k_{\rm B}T}

\newcommand{\cref}{c^{(2)}_{\rm ref}}

\begin{document}
\title{Monolayers of hard rods on planar substrates: I. Equilibrium}

\author{M. Oettel$^{1}$, M. Klopotek$^1$, M. Dixit$^2$, E. Empting$^1$, T. Schilling$^2$, and H. Hansen--Goos$^3$}
\affiliation{ 
$^1$ Institut f\"ur Angewandte Physik, Eberhard Karls Universit\"at T\"ubingen, D--72076 T\"ubingen, Germany \\
$^2$ Universit\'e du Luxembourg, Theory of Soft Condensed Matter, Physics and Materials Sciences Research Unit, L-1511 Luxembourg, Luxembourg \\ 
$^3$ Institut f\"ur Theoretische Physik, Eberhard Karls Universit\"at T\"ubingen, D--72076 T\"ubingen, Germany
}

\email{martin.oettel@uni-tuebingen.de}

\begin{abstract}
  The equilibrium properties of hard rod monolayers are investigated in a lattice model (where position and orientation of a rod
are restricted to discrete values) as well as in an off--lattice model featuring spherocylinders with continuous positional and orientational degrees of freedom.
Both models are treated using density functional theory and Monte Carlo simulations. Upon increasing the density of rods in the
monolayer, there is a continuous ordering of the rods along the monolayer normal (``standing up" transition). The continuous transition
also persists in the case of an external potential which favors flat--lying rods in the monolayer. This behavior is 
found in both the lattice and the continuum model. For the lattice model, we find very good agreement between
the results from the specific DFT used (lattice fundamental measure theory) and simulations. The properties
of lattice fundamental measure theory are further illustrated by the phase diagrams of bulk hard rods in two and three dimensions.    
\end{abstract}

\maketitle

\section{Introduction}

Several systems of scientific and technological interest can be characterized as
being monolayers of anisotropic particles, such as Langmuir monolayers \cite{review-Kaganer-Mohwald-Dutta}
or very thin films of elongated organic molecules \cite{review-Schreiber} such as organic 
semiconductors \cite{review-OMBD-Schreiber,review-OMBD-Witte-Woell}.
Since the particular molecular interactions in these systems may be very complicated,
it is worthwhile to investigate simpler models of anisotropic colloids to obtain general insights
into the thermal behavior of these systems. 
Among these, hard--body models (where particles interact only via their excluded volume) are
a natural starting point to assess effects of anisotropy, both for thermal equilibrium and
non--equilibrium conditions (i.e. growth of the monolayer).

We present our investigations on equilibrium and on growth of hard--rod monolayers in two papers, where
in the present first one, we focus on equilibrium properties and in the second, we treat the growth
process. In both papers, the focus will be on hard--rod lattice models since for these a fairly transparent theoretical
analysis of equilibrium and growth is possible in the framework of density functional theory (DFT).
In particular, we use the framework provided by fundamental measure theory (FMT) \cite{Ros89}   
within which very accurate density functionals for systems of anisotropic hard particles have been 
constructed for continuous \cite{Han09} and, important for the present investigations, also for 
lattice models \cite{Laf02,Laf04}.
For the subsequent investigations of the growth process, kinetic Monte--Carlo simulations
on a lattice are a natural starting point.
Since lattice models inevitably restrict the translational and orientational degrees of freedom of rods, we will
also present results for an off--lattice hard rod (spherocylinder) model and identify similarities and differences
between lattice and off--lattice models.        

The restriction of orientation in hard--rod (cuboid) models comes with the benefit that density functionals from 
FMT become tractable and therefore also analytic results can be derived.
For continuous translational degrees of freedom and in three dimensions (3d) a rich phase
diagram was derived \cite{Mar04}. Although not all details are the same in a simulated phase diagram
of hard cuboids with unrestricted orientation \cite{Esc08}, the restricted orientation model gives a
good first estimate of what can be expected. 
If the particles are restricted to a plane (the monolayer case), the first--order isotropic--nematic
transition becomes a continuous one, according to FMT in the restricted orientation model \cite{Mar14,Mar15}. The
orientational order perpendicular to the plane $Q$ is proportional to the density $\rho$ for low densities.
An approximate DFT and simulations for hard ellipsoids (unrestricted orientation) seem to confirm this behavior although 
very low densities have not been sampled in the simulations \cite{Var16}. Such a possible qualitative change of the nature of
the nematic transition through dimensional restriction is very interesting by itself, and therefore we will establish
analytically the $Q \propto \rho$ behavior explicitly in the low--density limit for both lattice and
continuum models. The presence of an orientation--dependent external potential in the monolayer plane (substrate potential) does
not change the continuous nature of the nematic transition but the onset of particles ``standing up'' may become
very sharp for substrate potentials which actually favor particles ``lying down''.        

The structure of the paper is as follows: In Sec. \ref{sec:latticedft}, 
we describe the lattice version of fundamental measure theory (FMT) for hard rod mixtures and give
illustrative examples for the functionals.  
The bulk equilibrium properties of monocomponent rods in two dimensions (2D) and three dimensions (3D) are 
briefly discussed, followed by the results for the monolayer (3D confined). 
Sec.~\ref{sec:offlattice} discusses the spherocylinder off--lattice model for the monolayer using DFT in the
low--density limit and using simulations.  Sec.~\ref{sec:summary} 
discusses similarities and differences between the lattice and off--lattice models and gives a summary. 
Two appendices briefly discuss the grand canonical simulation method for the lattice model and the derivation of
the excluded area between hard rods (in the continuum model) whose centers are confined to a plane. 

\section{Density Functional Theory for hard rod lattice models}
\label{sec:latticedft}

\subsection{Fundamental measure theory}

\begin{figure}
 \epsfig{file=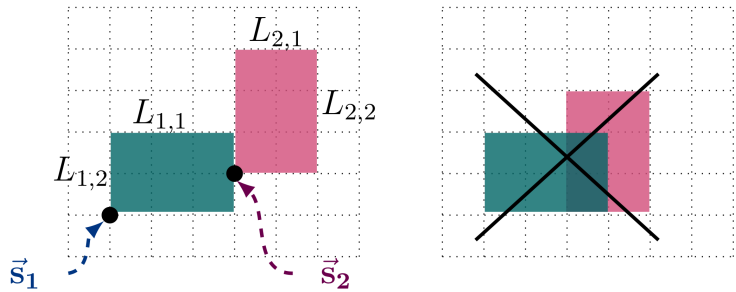, width=10cm}
 \caption{(color online) Definitions for the example of hard 2x3--rods in $d=2$. Rod location is specified by the position of the lower left corner
 (i.e., the corner whose lattice coordinates are minimal). Rods may ``touch'' (left picture) but not overlap (right).}
 \label{fig:rod_def}
\end{figure}

The rod model used in this work is formulated on a simple cubic lattice in $d$ dimensions. A lattice point $\vect s$ 
is specified by a set of $d$ integers ($\vect s=(s_1,...,s_d)$). The lattice constant $a$ is the unit of length. Hard rods
are lines (1D), rectangles (2D) or parallelepipeds (3D) with corners sitting on lattice points
and thus their geometry is
specified by their extent in the cartesian directions which are again sets of $d$ integers. 
The position of a rod is specified by the corner whose lattice coordinates are minimal each (see Fig.~\ref{fig:rod_def}).
Hard rods are not allowed to overlap (but they may ``touch'', i.e. share surfaces), 
thus the interaction potential for two rods ${\cal L}_i$ and  ${\cal L}_j$ of species
$i$ and $j$ at positions $\vect s_i$ and $\vect s_j$ with extensions
$\vect L_i=(L_{i,1},...,L_{i,d})$ and $\vect L_j=(L_{j,1},...,L_{j,d})$ is given by
\bea
 u_{ij}(\vect s_i,\vect s_j) &=& \left\{  \begin{matrix} \infty & \qquad (f_{ij}=1) \\ 
                                         0 & \qquad (f_{ij}=0)   \end{matrix} \right. 
\eea 
Here, $f_{ij}=f(\vect s_i,\vect s_j,\vect L_i, \vect L_j)$ is the rod overlap function given by
\bea
 \label{eq:foverlap}
 f(\vect s_i,\vect s_j,\vect L_i, \vect L_j) &= & \prod_{k=1}^d \theta(s_{i,k},s_{j,k},L_{i,k},L_{j,k}) \\
 \label{eq:theta} 
           \theta(s_{i,k},s_{j,k},L_{i,k},L_{j,k}) &=& \left\{ 
           \begin{matrix} 1 & \qquad (s_{j,k} = \{s_{i,k}-(L_{j,k}-1), ..., s_{i,k}+(L_{i,k}-1) \} ) \\
                          0 & \qquad (\mbox{otherwise})  \end{matrix} \right. 
\eea
The overlap function is 1 whenever there is 
overlap in all lattice dimensions, meaning that the rods are disjunct for $f=0$ (see. Fig.~\ref{fig:rod_def}). 
Note that due
to the chosen convention for the rod location the overlap function is not symmetric in the rod locations
$\vect s_i$ and $\vect s_j$. 

In the following, we consider such a rod mixture with $\nu$ species subject to  
external fields $V^{\rm ext}(\vect s)=\{V_1^{\rm ext}(\vect s),...V_\nu^{\rm ext}(\vect s)\}$ 
where $V_j^{\rm ext}(\vect s)$ acts on rod species $j$. At lattice site $\vect s$, the number density of rods per 
lattice site is specified
by $\rho(\vect s)=\{\rho_1(\vect s),...,\rho_\nu(\vect s)\}$ where $\rho_j(\vect s)$ is the
density of rod species $j$, i.e. the probability of a given site to be occupied by the lower left 
corner of a particle.
In density functional theory, all equilibrium properties of a rod mixture in external fields 
are obtained by minimizing the grand potential functional
\bea
  \Omega[\rho(\vect s)] &=& {\cal F^{\rm id}}[\rho(\vect s)] + {\cal F^{\rm ex}}[\rho(\vect s)] - 
   \sum_{i=1}^\nu \sum_{\vect s} (\mu_i - V^{\rm ext}_i(\vect s)) \, \rho_i(\vect s)
\eea 
with respect to the particle densities $\rho(\vect s)$.
%Here, $\mu=\{\mu_1,...,\mu_\nu\}$ is a vector with components which are the chemical potentials for the $\nu$ different 
%rod species and the dot product
%acts between the $\nu$--dimensional vectors in species space.
The chemical potential for rod species $i=1...\nu$ is denoted by $\mu_i$ 
If different species belong to the same type of rod in different orientations,
the corresponding chemical potentials must be equal in equilibrium. 
 ${\cal F^{\rm id}}[\rho(\vect s)]$ denotes the ideal gas contribution to
the free energy functional, given by 
\bea
  {\cal F^{\rm id}}[\rho(\vect s)] &=& \sum_{i=1}^\nu \sum_{\vect s}  \rho_i(\vect s) (\ln \rho_i(\vect s) - 1) \;.
\eea 
Energies are measured in units of $\kt$ throughout the paper.
%Here, $\beta =1 /\kt$ is the inverse temperature. 
%and $\vect 1$ is a $\nu$--dimensional vector with entry 1 in each component.

The exact form of the excess free energy functional ${\cal F^{\rm ex}}$ is in general unknown, in this work we will approximate it 
within the fundamental measure approach. For lattice models of hard rods, this approach has been worked out in 
Refs.~\cite{Laf02,Laf04}, resulting in an approximative form for ${\cal F^{\rm ex}}$ which we apply in the present study
(Lafuente--Cuesta functional). 
%In App. \ref{app:fmt} we will sketch the main idea which is based on dimensional reduction, and we provide some examples
%which might be useful for the interested reader.

\begin{figure}
 \epsfig{file=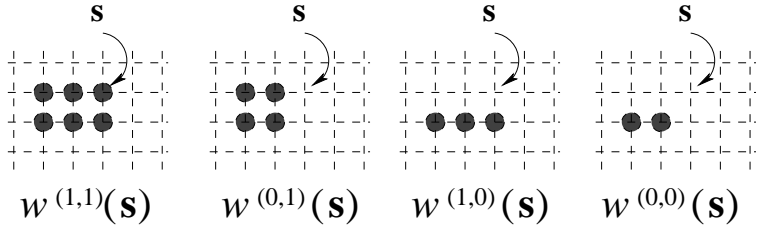, width=10cm}
 \caption{The four FMT weight functions for a rod with edge lengths $\vect L=(3,2)$. The lattice point
 at which the weight functions are evaluated 
 is denoted by $\vect s$. The thick points indicate on which lattice points the weight function is 1.}
 \label{fig:weights}
\end{figure}   

The class of free energy functionals derived in Refs.~\cite{Laf02,Laf04} makes use of weighted densities $n^\alpha(\vect s)$ which
are defined as convolutions of densities $\rho(\vect s)$ with weight functions
$w^\alpha := \{w_1^\alpha, ..., w_\nu^\alpha\}$:
\bea
   n^\alpha(\vect s) &= & \sum_{i=1}^\nu \rho_i \otimes  w^\alpha_i\,(\vect s) % =: \rho \otimes w^\alpha (\vect s) \;.
\eea
%As can be seen, $\otimes$ is a dot product in species space which sums over 
Convolutions ($\otimes$) on the lattice are defined as
\bea
   (f \otimes g) \, (\vect s) &=& \sum_{\vect s'} f(\vect s) g(\vect s' - \vect s) \;.
\eea
The $d$--dimensional index $\alpha=(\alpha_1,...\alpha_d)$ specifies different weight functions
$w_i^{\alpha}$, with allowed values $\alpha_i={0,1}$ only.
The weight functions $w_i^{\alpha}$ (specific for species $i$) have the meaning of defining a support of rods ${\cal K}_i^\alpha$ with edge lengths 
$\vect K_i^\alpha=(K_{i,1}^{\alpha_1},...,K_{i,d}^{\alpha_d})$, i.e. they are 1 on points covered by ${\cal K}_i^\alpha$
and 0 otherwise. This can be formalized using the $\theta$--function already employed for defining
rod overlap (Eq.~(\ref{eq:theta})),
\bea
  w_i^{\alpha}(\vect s) &=&  \prod_{k=1}^d \theta(0,s_{k},\alpha_k,K^{1}_{i,k}) \;. 
\eea
The edge lengths of rods ${\cal K}_i^\alpha$ are related to those of the rods ${\cal L}_i$ as follows:
\bea
 K_{i,k}^{\alpha_k} = L_{i,k} - (1 -\alpha_k) \qquad (k=1,...,d)\;,
\eea
i.e. whenever the index $\alpha_j$ is 0, the edge length of
${\cal K}_i^{\alpha_i}$ in the dimension $j$ is shortened by 1 compared to the corresponding edge length of ${\cal L}_i$,
otherwise ($\alpha_j=1$) the edge length is identical. 
In particular, for $\alpha =(1,...,1)$
all rods ${\cal K}_i^{\alpha_i}$ are identical to ${\cal L}_i$. The corresponding weighted densitiy
$n^{(1,...,1)}(\vect s)=\eta(\vect s)$ has the meaning of a local packing or volume fraction of rods at point $\vect s$.  
Fig.~\ref{fig:weights} illustrates the four possible weight functions for a rods with edge lengths $\vect L=(3,2)$
on a 2D lattice.

As a second ingredient, the Lafuente--Cuesta functional  
needs the excess free energy of a zero--dimensional (0d) cavity, $\Phi^{0d}$, i.e. a restricted domain on the 
lattice which can only hold one particle at a time.  
Such a cavity may consist of more than one point where the rod is positioned. Furthermore,
for a mixture the set of points $\{s_{{\rm cav},i}\}$ specifying the allowed location of species $i$ does not need to coincide
with the corresponding set $\{s_{{\rm cav},j}\}$ for species $j$. 
Note that the sets $\{s_{{\rm cav},i}\}$ and $\{s_{{\rm cav},j}\}$ are not independent since the 0D cavity property is required to hold globally for the mixture and not just for the individual components.
%Examples are discussed in App.~\ref{app:fmt}.
The  free energy $\Phi^{0d}(\eta)$ of such a cavity is a function only of the total packing fraction
$\eta \equiv \eta_{\rm cav} = \sum_{i=1}^\nu \sum_{\vect s \in \lbrace s_{{\rm cav},i} \rbrace } \rho_i(\vect s)$
in the cavity,
\bea
 \label{eq:f0d}
 \Phi^{0d}(\eta) &=& \eta + (1-\eta) \ln(1-\eta) \;.  
\eea
Using this 0d free energy, the  Lafuente--Cuesta excess free energy functional is given by
\bea
 \label{eq:fex}
 {\cal F^{\rm ex}} &=& \sum_{\vect s} {\cal D}_\alpha \Phi^{0d}(n^\alpha(\vect s)) \;.  
\eea 
Remember that $\alpha$ is a $d$--dimensional index with entries $\{ 0,1 \}$ only.
In Eq.~(\ref{eq:fex}), ${\cal D}_\alpha = \prod_{i=1}^d D_{\alpha_i}$ and $D_{\alpha_i}$ is the difference operator whose action
on a function $f(\alpha_i)$ is given by $D_{\alpha_i} f(\alpha_i) = f(1)-f(0)$. 

It can be shown that ${\cal F^{\rm ex}}$ as defined above yields the correct excess free energy, 
Eq.~(\ref{eq:f0d}), for {\em any} 0D cavity \cite{Laf02,Laf04}. In order to assess the accuracy of the 
expression for situations of less severe confinement, we evaluate explicitly the properties of 
different bulk systems in Sec. \ref{sec:bulk2d3d}. In a first step, however, we illustrate the
construction of the Lafuente--Cuesta functional by applying it to different mixtures in 1D, 2D, and 3D.

\subsection{Special cases}

Here we give the explicit functionals for some special mixtures. The equilibrium properties of examples (b) and (c) 
(2D and 3D systems) will be discussed in Sec.~\ref{sec:bulk2d3d} and those of example (d) (monolayer) in
Sec.~\ref{sec:monolayer}.
\begin{itemize}
 \item[(a)] $d=1$: Mixture of hard rods in one dimension. The excess free energy functional is given by
\bea
 \label{eq:fex_1d}
 {\cal F^{\rm ex}} &=& \sum_{\vect s} \left(\Phi^{0d}(n^{(1)}(\vect s))-\Phi^{0d}(n^{(0)}(\vect s)) \right) \;.  
\eea  
This is the well--known exact solution for the 1d lattice hard rod mixture, derived in Ref.~\cite{Laf02} following the recipe from 
Ref.~\cite{Per89} which treats the 1D continuum hard rod mixture. 
Another yet different derivation can be found in Ref.~\cite{Maass12}. 

 \item[(b)] $d=2$: A system of rods with length $L$ and width 1 corresponds to the binary mixture with rod lengths
   $\vect L_1=(L,1)$ and $\vect L_2=(1,L)$.  The excess free energy functional is given by
\bea
 \label{eq:fex_2d}
  {\cal F^{\rm ex}} &=& \sum_{\vect s} \left(\Phi^{0d}(n^{(1,1)}(\vect s))-\Phi^{0d}(n^{(0,1)}(\vect s))-
  \Phi^{0d}(n^{(1,0)}(\vect s)) \right) \;.  
\eea  
 The weighted densities are given by 
\bea
 n^{(1,1)}(\vect s) &=& \rho_1 \otimes w_1^{(1,1)}\,(\vect s) + \rho_2 \otimes w_2^{(1,1)}\,(\vect s) \;, \nonumber \\
 n^{(0,1)}(\vect s) &=& \rho_1 \otimes w_1^{(0,1)}\,(\vect s) \;, \\
 n^{(1,0)}(\vect s) &=& \rho_2 \otimes w_2^{(1,0)}\,(\vect s) \;. \nonumber 
\eea
 Note that the weights $w_2^{(0,1)}=w_1^{(1,0)} =0$ since they correspond to the support of rods with width 0.
 Likewise $w_1^{(0,0)}=w_2^{(0,0)} =0$.
  
 \item[(c)] $d=3$: A system of rods with length $L$ and height/width 1 corresponds to the ternary mixture with rod lengths
   $\vect L_1=(L,1,1)$, $\vect L_2=(1,L,1)$ and $\vect L_3=(1,1,L)$.  The excess free energy functional is given by
\bea
 \label{eq:fex_3d}
  {\cal F^{\rm ex}} &=& \sum_{\vect s} \left(\Phi^{0d}(n^{(1,1,1)}(\vect s))-\Phi^{0d}(n^{(0,1,1)}(\vect s))-
  \Phi^{0d}(n^{(1,0,1)}(\vect s)) - \Phi^{0d}(n^{(1,1,0)}(\vect s)) \right)  \;. \qquad  
\eea  
 The weighted densities are given by 
\bea
 n^{(1,1,1)}(\vect s) &=& \rho_1 \otimes w_1^{(1,1,1)}\,(\vect s) + \rho_2 \otimes w_2^{(1,1,1)}\,(\vect s) + \rho_3 \otimes w_3^{(1,1,1)}\,(\vect s)\;, \\
 n^{(0,1,1)}(\vect s) &=& \rho_1 \otimes w_1^{(0,1,1)}\,(\vect s) \;, \nonumber \\
 n^{(1,0,1)}(\vect s) &=& \rho_2 \otimes w_2^{(1,0,1)}\,(\vect s) \;, \\
 n^{(1,1,0)}(\vect s) &=& \rho_3 \otimes w_3^{(1,1,0)}\,(\vect s) \;. \nonumber 
\eea
 Similarly to case (b), the weights $w_i^{(\alpha_1,\alpha_2,\alpha_3)}=0$ whenever $\alpha_j=0$ and $i \neq j$ 
 since they correspond to the support of rods with width 0.

 \item[(d)] $d=3$ (confined), the monolayer: A system of rods with length $L$ and height/width 1 whose positions are constrained 
 to a 2D--plane corresponds to a 2D ternary mixture with rod lengths
   $\vect L_1=(L,1)$, $\vect L_2=(1,L)$ (rods lying in--plane) and $\vect L_3=(1,1)$ (rods standing up).  
   The excess free energy functional is given by formally the same functional as in (a),
\bea
 \label{eq:fex_2+1d}
  {\cal F^{\rm ex}} &=& \sum_{\vect s} \left(\Phi^{0d}(n^{(1,1)}(\vect s))-\Phi^{0d}(n^{(0,1)}(\vect s))-
  \Phi^{0d}(n^{(1,0)}(\vect s)) \right) \;,  
\eea  
 but now the weighted densities are given by 
\bea
 n^{(1,1)}(\vect s) &=& \rho_1 \otimes w_1^{(1,1)}\,(\vect s) + \rho_2 \otimes w_2^{(1,1)}\,(\vect s) + \rho_3 \otimes w_3^{(1,1)}\,(\vect s)\;, \nonumber \\
 n^{(0,1)}(\vect s) &=& \rho_1 \otimes w_1^{(0,1)}\,(\vect s) \;, \\
 n^{(1,0)}(\vect s) &=& \rho_2 \otimes w_2^{(1,0)}\,(\vect s) \;. \nonumber 
\eea

\end{itemize}

\subsection{Equilibrium bulk properties in 2D and 3D}
\label{sec:bulk2d3d}

\begin{figure}
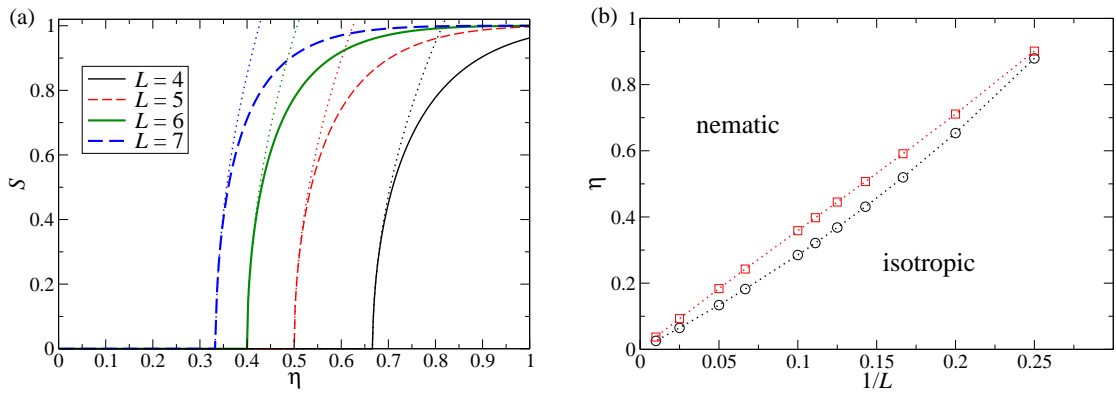

  \epsfig{file=fig3a.eps, width=7cm} \hspace{5mm}
  \epsfig{file=fig3b.eps, width=7cm} 
  \caption{(color online) (a) Rods in $d=2$: Demixing order parameter $S$ as a function of the total packing fraction for different rod lengths $L$.
   Dotted lines correspond to the approximate solution near the onset of demxing (Eq.~(\ref{eq:2dSapprox})).  }
   (b) Rods in $d=3$. Liquid--nematic binodal in the plane spanned by the inverse rod length $1/L$ and the packing fraction $\eta$. Square
   symbols show the packing fraction of the coexisting nematic state, circles the packing fraction of the coexisting liquid state.
  \label{fig:2d_eq}
\end{figure}

\subsubsection{$d=2$, the binary mixture with rod lengths $\vect L_1=(L,1)$ and $\vect L_2=(1,L)$}

In the bulk, both densities ($\rho_1$ and $\rho_2$) and all weighted densities are constant.
We introduce the total density $\rho:=\rho_1+\rho_2$ and
denote by $\eta := n^{(1,1)}=L \rho$ the total packing fraction. Furthermore
$n^{(0,1)}=(L-1)\rho_1$, $n^{(1,0)}=(L-1)\rho_2$ and $S=(\rho_1-\rho_2)/\rho$ is an order parameter
for the demixed state. We refrain from calling $S$ a nematic order parameter, since the alignment
of rods corresponds just to a demixed state between species 1 and 2, and the corresponding transition has
the character of a liquid--vapor transition \cite{Vink09}. The bulk free energy density, 
$f_{2d}(\rho,S)=f^{\rm id}_{2d}+f^{\rm ex}_{2d}$,
written in dependence on the variables $\rho$ and $S$ becomes
\bea
 f^{\rm id}_{2d} &=& \sum_{i=1}^3\rho_i \ln \rho_i  - \rho \;,   \\
%\frac{\rho}{2} \left( (1+S) \ln\left[ \frac{\rho}{2}(1+S)\right] + (1-S) \ln\left[ \frac{\rho}{2}(1-S)\right] - 2   \right)  \\
 f^{\rm ex}_{2d} &=&  \Phi^{0d}(\rho L) -\Phi^{0d} \left( (L-1)\rho_1 \right) -
                                         \Phi^{0d} \left(  (L-1)\rho_2  \right) \;,  \\
    \rho_{1} &=& \frac{\rho}{2}(1 + S) \;,  \\
    \rho_{2} &=& \frac{\rho}{2}(1 - S) \;. 
\eea 
At fixed $\rho$, the equilibrium demixing parameter $S_{\rm eq}$ is found by solving 
$\mu_S=\partial f_{2d}/\partial S=0$. For $L\le3$, the mixed state ($S_{\rm eq}=0$)
is the only solution and $f$ is minimal there, For $L\ge 4$ there exists a critical packing
fraction $\eta_c <1 $ above which three solutions $S=\{0,\pm S_{\rm eq}\}$ signal demixing:
the solutions $S \neq 0$ have lower free energy.
At $\eta_c$, there is no jump in the demixing parameter which is the behavior also observed at a 
liquid--vapor transition. Therefore
one may expand
\bea
  \mu_S(\eta,S) \approx \mu_{1,S}(\eta) S + \mu_{3,S}(\eta) S^3 + \dots 
  \label{eq:muapprox}
\eea
and find the critical packing fraction by solving $\mu_{1,S}(\eta_c)=0$, with the solution
\bea
  \eta_c = \frac{2}{L-1} \;.
\eea
The equilibrium demixing $S_{\rm eq}(\eta)$ in the vicinity of $\eta_c$ can be approximated 
by solving $\mu_S=0$ for $S$ using the Taylor approximation (\ref{eq:muapprox}), giving
\bea
  S_{\rm eq} = \sqrt{ - \frac{\mu_{1,S}(\eta)}{\mu_{3,S}(\eta)}} \approx \sqrt{\eta-\eta_c} 
         \,\sqrt{\frac{3}{2(L-2)}} (L-1) \;.
 \label{eq:2dSapprox}
\eea
The behavior of  $S_{\rm eq}(\eta)$ near $\eta_c$ born out by the approximate theory is of course of mean-field type.

These findings can be compared with simulation work which finds the demixing transition
for $L \ge 7$ \cite{Ghosh07} and a critical packing fraction $\eta_c \approx 5/L$ \cite{Mat08}.
Thus, FMT overestimates the tendency to demix. Note, however, that the demixing
follows from a {\em single} functional, unlike other approaches which assume distinct 
epxressions for the isotropic and the demixed phase free energies \cite{Mat08}.

\subsubsection{$d=3$, the ternary mixture with rod lengths
   $\vect L_1=(L,1,1)$, $\vect L_2=(1,L,1)$ and $\vect L_3=(1,1,L)$}  

The total density is $\rho=\rho_1+\rho_2+\rho_3$ and the total packing fraction 
is $\eta:= n^{(1,1,1)}=L\rho$. We define the order parameters 
\bea
 Q & =& \frac{\rho_3-\frac{\rho_1+\rho_2}{2}}{\rho_1+\rho_2+\rho_3} \;, \nonumber \\
 S & = & \frac{\rho_1-\rho_2}{\rho_1+\rho_2}\;. \label{eq:orderparam3D}
\eea
$Q \not = 0$ signifies an excess ($Q>0$) or depletion ($Q<0$) of particles in $z$--direction (nematic state) while $S \neq 0$ signals order
in the $x$--$y$--plane orthogonal to the nematic director (biaxial state).
The bulk free energy density, 
$f_{3d}(\rho,Q,S)=f^{\rm id}_{3d}+f^{\rm ex}_{3d}$,
written in dependence on the variables $\rho$, $Q$ and $S$ becomes
\bea
 \label{eq:fid_3d_bulk}
 f^{\rm id}_{3d} &=&  \sum_{i=1}^3\rho_i \ln \rho_i  - \rho \;,   \\
 \label{eq:fex_3d_bulk}
 f^{\rm ex}_{3d} &=&  \Phi^{0d}(L\rho) -\Phi^{0d} \left( (L-1)\rho_1 \right) -
                                         \Phi^{0d} \left( (L-1)\rho_2 \right) - \Phi^{0d} \left( (L-1)\rho_3  \right) \;, \\
    \rho_{1} &=& \frac{\rho}{3}(1-Q)(1 + S) \;, \\ 
    \rho_{2} &=& \frac{\rho}{3}(1-Q)(1 - S) \;, \\ 
    \rho_3      &=& \frac{\rho}{3}(1+2Q) \;.   
\eea 
Minimization of the total free energy density with respect to $Q$ and $S$ shows that the model
has a stable nematic state ($Q=Q_{\rm min}>0$, $S=0$) for $L \ge 4$. Note that 
the director could also be oriented along the $x$-- or $y$--axis instead of the chosen $z$--axis. 
A pure nematic state with director along the $x[y]$--axis and order parameter $Q'$ is equivalent to
a minimum free energy state with $Q=-Q'/2$ and $S=\pm 3Q'/(2+Q')$ using the order parameters 
(\ref{eq:orderparam3D}). Therefore this is not a biaxial state. The associated liquid--nematic transition is
of first order, and we have determined coexistence between the liquid and the nematic state by performing
the common tangent construction for the free energy density $f_{3d}(\rho,0,0)$ (liquid phase) and
$f_{3d}(\rho,Q_{\rm min},0)$ (nematic phase) which implies equality of the chemical potential 
$\mu = (\partial f_{3d})/(\partial\rho)$ and pressure $p = \mu\rho-f_{3d}$. Results are shown in 
Fig.~\ref{fig:2d_eq}(b). The packing fractions of the coexisting nematic state are very well described
by $\eta_{\rm c,nem}=3.58/L$, and the gap in packing fractions of the coexisting states has a maximum
of $\approx 0.08$ at $L=8$ and tends to zero as $L \to \infty$.

\begin{figure}
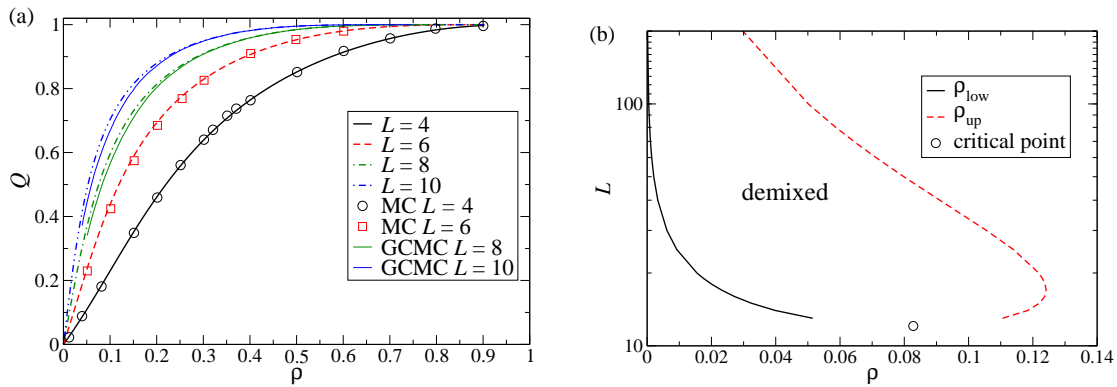

  \epsfig{file=fig4a.eps, width=7cm} \hspace{5mm}
  \epsfig{file=fig4b.eps, width=7cm}
  \caption{(color online) (a) Order parameter $Q$ for rods standing up vs. total density. Lines are results from FMT and symbols are
  results from Monte Carlo simulations reported in Ref.~\cite{Kra92}. {Thin lines are results from our GCMC simulations
  where a running average of 20 points on density intervals of 0.04 has been taken.} (b) Phase diagram {from FMT} showing a reentrant behavior for mixing
   ($S=0$) and demixing ($S \neq 0$) in the plane. The rod length $L$ is treated as a continuous variable. 
  The critical point occurs for a rod length of $L_c \approx 12.077$ at a
  density of $\rho_c \approx 0.0828$. }
  \label{fig:2d3d_eq}
\end{figure}

\subsection{The monolayer system}
\label{sec:monolayer}

In the lattice model, this is the effectively 2D ternary mixture with rod lengths
$\vect L_1=(L,1)$, $\vect L_2=(1,L)$ (rods lying in--plane) and $\vect L_3=(1,1)$ (rods standing up).  

The total density is $\rho=\rho_1+\rho_2+\rho_3$ and the total packing fraction in the plane
is $\eta:= n^{(1,1)}=L(\rho_1+\rho_2)+\rho_3$. The order parameters $Q$ and $S$ are the same as in Eqs.~(\ref{eq:orderparam3D}).
$Q > 0$ signifies an excess of particles ``standing--up'' (nematic state) while $S \neq0$ signals demixing
of ``lying--down'' particles (biaxial state, if additionally $Q\neq0$).
In the bulk free energy density, 
$f_{3d,{\rm conf}}(\rho,Q,S)=f^{\rm id}_{3d,{\rm conf}}+f^{\rm ex}_{3d,{\rm conf}}$,
one can identify $f^{\rm id}_{3d,{\rm conf}}=f^{\rm id}_{3d}$ whereas the excess part becomes
\bea
 f^{\rm ex}_{3d,{\rm conf}} &=&  \Phi^{0d}(L(\rho_1+\rho_2)+\rho_3) -\Phi^{0d} \left( (L-1)\rho_1\right) -
                                         \Phi^{0d} \left( (L-1)\rho_2  \right) \;. 
 \label{eq:fex_mono_bulk}
\eea 
At fixed total density $\rho$, the minimization of the free energy with respect to $Q$ and $S$ gives the following picture.
For ``small'' rod lengths $L \le 12$ there is no biaxial state ($S=0$, no demixing in the plane) but the ``nematic'' order
parameter $Q$ monotonically and smoothly grows from 0 to 1 when the total density varies between 0 and 1 (close-packed state of
rods standing up). Results for $L=4 ... 10$ are shown in Fig.~\ref{fig:2d3d_eq}(a) which demonstrates that for increasing $L$
the rods 
quickly ``stand up''. The FMT results show excellent agreement with Monte Carlo simulation results \cite{Kra92}
on the same confined model for $L=4$ and 6.
{For larger rod lengths ($L=8$ and 10) the agreement with our grand canonical Monte Carlo (GCMC) simulations is
only slightly worse. The implementation of GCMC is briefly described in App.~\ref{app:gcmc}.} 

For $L \ge 13$, FMT predicts reentrant demixing in the plane, i.e. in a certain 
interval $[\rho_{\rm low}(L), \rho_{\rm up}(L)]$ for the total density the biaxiality parameter will be nonzero,  $S \neq 0$.
This reentrant behavior is qualitatively understood as follows. In the $d=2$ model it was found that the critical density of demixing 
of planar rods is $\rho_1+\rho_2 = 2/(L(L-1))$. For increasing $L$ one therefore expects $\rho_{\rm low}(L) \to 0$. On the other hand,
for a certain  $L$ but increasing $\rho$  the fraction of planar rods initially grows, reaches a maximum and becomes smaller again 
since the rods stand up, see
Fig.~\ref{fig:2d3d_eq}(a). Therefore, if there exists a lower demixing density $\rho_{\rm low}(L)$ then one would expect the existence of 
a higher remixing density $\rho_{\rm up}(L)$ owing to the reduction of the planar rod density.      
As in the $d=2$ model, the demixing transition is continuous and therefore the densities $\rho_{\rm low}(L), \rho_{\rm up}(L)$
can be found by the following argument: Let $\mu_Q(\rho,Q,S)=\partial f_{3d,{\rm conf}}/\partial Q$ and 
$\mu_S(\rho,Q,S)=\partial f_{3d,{\rm conf}}/\partial S$ be chemical potentials for the order parameters $Q$ and $S$.
For a mixed state ($S=0$), we define $Q_{\rm eq} (\rho)$ through $\mu_Q(\rho,Q_{\rm eq},0)=0$. As before, we may expand 
\bea
  \mu_S(\rho,Q,S) \approx \mu_{1,S}(\rho,Q) S + \mu_{3,S}(\rho,Q) S^3 + \dots 
  \label{eq:muapprox1}
\eea
At the de-/remixing densities one has the condition 
\bea
 \mu_{1,S}(\rho,Q_{\rm eq}(\rho))|_{\rho=\rho_{\rm low[up]}}  = 0 ,
\eea 
which needs to be solved numerically. The results are shown in Fig.~\ref{fig:2d3d_eq}(b), showing the onset of demixing at $L=13$ and a maximum
density interval for the demixed state at around $L=20$. 

The continuous behavior of $Q(\rho)$ and the reentrant demixing are actually very similar to the behavior found
in the FMT study of the restricted orientation model with continuous translational degrees of freedom \cite{Mar14}.
There biaxial ordering sets in at larger rod lengths, $L \ge 21.34$. 

The results in Fig.~\ref{fig:2d3d_eq} suggest $Q \propto \rho$, i.e. the continuous nematic ordering sets in at $\rho=0$.
This is easily understood in a low--density expansion of the FMT excess free energy (\ref{eq:fex_mono_bulk}) which is exact
up to second order. Assuming no biaxiality ($S=0$) and combining ideal and excess part we find for the
free energy derivative with respect to $Q$:
\bea
 \label{eq:lat_virial}
  \mu_Q = \frac{\partial f_{3d,{\rm conf}}}{\partial Q}  \approx \frac{2}{3} \rho \ln\frac{1+2Q}{1-Q} +  
           \frac{2}{9} \rho^2 \left( [2-L-L^2] + [L-1]^2 Q   \right) + \mathcal{O}(\rho^3) \;.
\eea  
Note that in the excess part of $\mu_Q$, at fixed density, there is a constant term driving the system to $Q>0$ for $L \ge 2$.
This is different from the 2D and 3D bulk systems where this constant term is absent and thus $Q>0$ (for low densities) is always 
unfavorable in terms of free energy cost. The equilibrium solution $\mu_Q=0$ at $Q=Q_{\rm eq}$ is found as
\bea
 \label{eq:Q_smallrho_lat}
 \rho = \frac{3 \ln\frac{1+2Q_{\rm eq}}{1-Q_{\rm eq}}}{[L^2+L-2] - [L-1]^2 Q_{\rm eq}}
  \quad \rightarrow \quad Q_{\rm eq} \approx \frac{1}{9}(L^2+L-2) \rho \;.
\eea  
Hence, for large $L$ the lattice model predicts a scaling $Q_{\rm eq} \propto \rho L^2$.

\begin{figure}
 \centerline{ 
  \epsfig{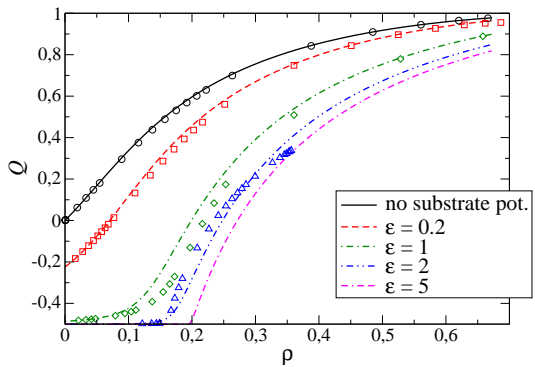}
 }
  \caption{(color online) Order parameter $Q$ for rods standing up vs. total density subject to a substrate potential (rod length $L=5$).
   The substrate potential is parametrized as $-\epsilon$ per unit length such that $v_3=-\epsilon$, $v_0=-L\epsilon$
   and thus $v_Q=(2/3)(L-1)\epsilon$. {Lines are DFT results, symbols results from GCMC simulations (see App.~\ref{app:gcmc}). 
  The error is smaller than the symbol size.}}
  \label{fig:2d3d_vext}
\end{figure}

\subsubsection{Finite substrate potential}

One may ask whether a finite substrate potential could alter the continuous transition found above. It is natural to assume
that the substrate potential acts equally on the flat--lying species 1 and 2 and differently on the upright species 3.
Hence the external contribution to the free energy becomes
\bea
 f^{\rm ext} = 
\sum_{i=1}^3 V^{\rm ext}_i \rho_i  = v_0(\rho_1+\rho_2) + v_3 \rho_3 
  = \frac{\rho}{3}(2v_0+v_3) + \frac{2}{3} \rho Q (v_3-v_0) \;. 
 \label{eq:fext_lat}
\eea 
Therefore the free energy derivative with respect to $Q$ is modified as $\mu_Q \to \mu_Q + \rho v_Q$ with 
$v_Q= \frac{2}{3} (v_3-v_0)$. For the ideal gas limit this implies an initial ordering on the substrate
with order parameter
\bea
   Q_{\rm id} = \frac{\exp(-3v_Q/2)-1}{\exp(-3v_Q/2)+2} \;.
\eea 
If the substrate is strongly attractive for the flat--lying species 1 and 2 ($v_Q \gg 0$), then  
we find $Q_{\rm id} \to - 1/2$. At nonzero densities, the solution of $\mu_Q=0$ (Eq.~\ref{eq:lat_virial} with
the external contribution) is obtained in the form $\rho(Q)$. For small deviations from equilibrium, $Q_{\rm eq}=Q_{\rm id} + \delta Q$,
we can invert this function and obtain
\bea
 \label{eq:deltaQ}
  \delta Q \approx \frac{3}{2}\rho \;  \frac{\alpha - \beta Q_{\rm id}}{ 2(1+2 Q_{\rm id})^{-1} + (1-Q_{\rm id})^{-1} }
\eea
with $\alpha=2(L^2+L-2)/9$ and $\beta=2(L-1)^2/9$. Although the range of validity is very limited, it implies that
the qualitative behavior for $\rho \to 0$ is unchanged since the slope of $\delta Q(\rho)$ is always positive. Thus the
transition stays continuous. However, for increasing $v_Q$ the ``standing up'' transition of the monolayer becomes
increasingly steep at {\em moderate} densities, see Fig.~\ref{fig:2d3d_vext} {where we show the $Q(\rho)$ behavior for $L=5$.
For these moderate densities the expansion up to second order is not valid anymore.
Especially for the case $\epsilon=5$ the behavior near $\rho=0.2$ looks as if $Q(\rho)$ has a bifurcation point here, similar to the
demixing transition in the 2D bulk system discussed in Sec.~\ref{sec:bulk2d3d}.
{The density $\rho=0.2=1/L$ at which this apparent transition occurs is the close--packing density for rods
lying flat.} 
However, {for finite potentials} it is not a phase transition} since $Q(\rho)$ maintains its linear behavior of $Q(\rho)$ with nonzero slope at very small densities.  

\section{Monolayers of hard spherocylinders}
\label{sec:offlattice}

For the lattice monolayer discussed in the previous section it does not matter which rod point or segment is actually fixed to the plane
since all choices lead to the same effective 2D model. Physically, fixing the end point corresponds to the case of rods on a hard substrate while
fixing some point in the middle of the rod applies to Langmuir monolayers. For a continuous model of hard rods, there should be a difference
between the two cases which is not expected to be qualitative (with regard to the type of transition). As can be seen below, the low--density behavior of long rods
with large aspect ratios
is actually insensitive to the choice of confining plane. Therefore we will present simulation results below only for the case of fixed mid--points. 

\subsection{DFT in an expansion up to second order in density}

We consider hard sperocylinders with length $L$ and diameter $D$
whose centers or ends are fixed on a plane.
In order to investigate the nature of the orientation transition, we consider a low--density expansion of the free energy.
For the well--studied model of hard rods in 3D, this method was used to establish the onset of nematic order as a bifurcation
and the nature as a first order transition \cite{Kay78}. 
The free energy density up to second order in density, including the contribution from an external potential, is given by
\bea
  {\cal F}  & = & {\cal F}^{\rm id} +  {\cal F}^{\rm ex} + {\cal F}^{\rm ext} \\
   {\cal F}^{\rm id} & =&   \int d^2 r \int d\Omega \; \rho(\vect r, \Omega) ( \ln (\rho(\vect r, \Omega) \Lambda^2) -1) \\
  \label{eq:fex_hardrods}
     {\cal F}^{\rm ex} & =& \frac{1}{2}  \int d^2 r \int d\Omega  \int d^2 r' \int d\Omega' \, \rho(\vect r, \Omega)
                              \rho(\vect r', \Omega') \omega ( |\vect r - \vect r'|, \Omega, \Omega') \\
     {\cal F}^{\rm ext} & =&   \int d^2 r \int d\Omega \; \rho(\vect r, \Omega) V^{\rm ext}(\vect r, \Omega)
\eea
Here, $\rho(\vect r, \Omega)$ is an inhomogeneous particle density in two dimensions and units of [length]$^{-2}$ which depends on the space point $\vect r$ and
the orientation of the rod $\Omega=(\theta,\phi)$, specified by the polar angle $\theta$ and the azimuthal
angle $\phi$. The integral over orientations is defined as
\bea
  \int d\Omega = \frac{1}{4\pi} \int_0^\pi \sin \theta d\theta \int_0^{2\pi} d\phi \;.
\eea
$\Lambda$ is the thermal de--Broglie length. $\omega(r,\Omega,\Omega')$ is the overlap function between
rods for given orientations of and distance $r$ between the particles.
It is 1 if there is overlap, otherwise zero. The external (substrate) potential $V^{\rm ext}(\vect r, \Omega)$ is measured
in units of $\kt$.

We consider only orientation--dependent substrate potentials, $V^{\rm ext}(\Omega)$,
and bulk states, i.e. no space dependence of the density and introduce the orientation distribution
$f(\Omega)$:
\bea
  \rho(\vect r, \Omega) = \rho_0 f(\Omega) \;.
\eea
Then the ideal, excess and external part of the  free energy per particle ($a= a^{\rm id} + a^{\rm ex} +  a^{\rm ext}$) become:
\bea
  a^{\rm id} &=& \int d\Omega f(\Omega) ( \ln(\rho_0 \Lambda^2 f(\Omega)) -1)\;, \\
  a^{\rm ex} &=& \frac{\rho_0}{2} \int d\Omega  \int d\Omega' \; f(\Omega) f(\Omega') \beta ( \Omega, \Omega')\;, \\
  a^{\rm ext} &=& \int d\Omega  \; f(\Omega) V^{\rm ext}(\Omega)\;. \label{eq:aext}
\eea
Here, $\beta ( \Omega, \Omega') = \int d^2 r \omega ( r, \Omega, \Omega')$ is the excluded area between the
rod centers (or ends) with fixed orientations of the rods.

In equilibrium, $f(\Omega)$ minimizes $a$. From $\delta a/\delta f=0$ we obtain
\bea
 \label{eq:feq}
  \ln f(\Omega) &=& - \ln C - V^{\rm ext}(\Omega) - \rho_0 \int d\Omega'\; \beta ( \Omega, \Omega') f(\Omega') \;,
\eea
where $C$ is a constant which must ensure that $f$ is properly normalized, i.e. $\int d\Omega\; f(\Omega)= 1$.
It is determined by exponentiating Eq.~(\ref{eq:feq}) and integrating over $\Omega$:
\bea
 \label{eq:feq2}
  f(\Omega) &=& \frac{1}{C} \exp\left( - V^{\rm ext}(\Omega)-\rho_0 \int d\Omega'\; \beta ( \Omega, \Omega') f(\Omega') \right) \;, \\
 \nonumber
       C &=& \int d\Omega \exp\left( - V^{\rm ext}(\Omega)-\rho_0 \int d\Omega'\; \beta ( \Omega, \Omega') f(\Omega') \right) \;.
\eea

The orientation--dependent substrate potential gives rise to a non--constant orientation distribution in the
ideal gas limit:
\bea
   f_{\rm id}(\Omega) = \frac{\exp(-V^{\rm ext}(\Omega))}{\int d\Omega \exp(-V^{\rm ext}(\Omega))}\;,
\eea  
which is normalized to 1.
We introduce the small deviation $f_1(\Omega):= f(\Omega)-f_{\rm id}(\Omega)$ and linearize Eq.~(\ref{eq:feq2}) in $f_1$:
\bea
  \label{eq:feqlin}
  \frac{f_1(\Omega)}{f_{\rm id}(\Omega)} &=& C_1  - \rho_0 \int d\Omega'  \beta ( \Omega, \Omega') (f_{\rm id}(\Omega') + f_1(\Omega')) \\
   \nonumber
       C_1 &=& \rho_0  \int d\Omega \int d\Omega' f_{\rm id}(\Omega) \beta ( \Omega, \Omega')  (f_{\rm id}(\Omega') + f_1(\Omega')) \;.
\eea
The constant $C_1$ ensures the necessary normalization condition $\int d\Omega f_1(\Omega) = 0$.
If one expands $f_1$ in powers of $\rho_0$ then one finds the leading order solution
\bea
 \label{eq:lo}
 f_1(\Omega) \approx \rho_0\; f_{\rm id}(\Omega)\left(    \int d\Omega \int d\Omega' f_{\rm id}(\Omega) \beta ( \Omega, \Omega')  f_{\rm id}(\Omega') 
      - \int d\Omega' \beta ( \Omega, \Omega')  f_{\rm id}(\Omega') \right) \;.
\eea
This expression is equivalent to Eq.~(\ref{eq:deltaQ}) in the lattice model and shows that any deviations from the
ideal gas distribution are continuous and proportional to the density $\rho_0$.

In the absence of a substrate potential ($f_{\rm id}=1$), we can proceed further.
Without loss of generality,
we put rod 1 at the coordinate center with orientation (director) $\vect u_1=(\sin\theta_1,0,\cos\theta_1)^T$. Rod 2 has the director
$\vect u_2=(\sin\theta_2 \cos\phi_2, \sin\theta_2 \sin\phi_2,\cos\theta_2)^T$. The excluded area depends in general
on the three angles $\theta_1,\theta_2,\phi_2$. 
If we consider only nematic order without biaxiality, $f(\Omega) \equiv f(\theta)$, 
then we can define the integrated
overlap area
\bea
  \frac{1}{2\pi} \int d\phi_2 \beta(\theta_1,\theta_2,\phi_2) =: \beta_\phi (\theta_1,\theta_2) \;.
\eea
If we take the polar angle (with respect to the interface normal) in the interval $[-\pi/2,\pi/2]$,
symmetry considerations give us
$\beta_\phi(\theta_1,\theta_2)=\beta_\phi(-\theta_1,\theta_2)=\beta_\phi(\theta_1,-\theta_2)=
\beta_\phi(-\theta_1,-\theta_2)$.
Since also $f(\theta) = f(-\theta)$, the integration domain over $\theta$ can be restricted to $[0,\pi/2]$. 
The nematic order parameter in the monolayer is defined by
\bea
 \label{eq:defS}
  Q_{\rm nem} = \int_0^{\pi/2} d(\cos \theta) P_2(\cos\theta) \; f(\theta) \;,
\eea
where $P_2(x)$ is the second of the Legendre polynomials $P_i(x)$. It is also useful to introduce the Legendre
coefficients of the excluded area:
\bea
   B_{ij} = \int_0^{\pi/2} d(\cos \theta) P_{2i}(\cos\theta) \int_0^{\pi/2} d(\cos \theta') P_{2j}(\cos\theta')\;
               \beta_\phi ( \theta, \theta') \;.
\eea
Owing to the symmetry of the excluded area, only projections onto even Legendre polynomials are nonzero.
Using these definitions, the nematic order parameter in the case of no substrate potential is obtained
by projecting with $P_2$ onto the solution for $f_1$ in Eq.~(\ref{eq:lo}):
\bea
 \label{eq:Qnem}
 Q_{\rm nem} \approx -\rho_0 B_{10}
\eea
This is an interesting result since it tells us that $Q_{\rm nem} \propto \rho_0$ as long as the leading off--diagonal
Legendre coefficient of the excluded area is nonzero. This is precisely the case in the monolayer system (see below), whereas
in 3D this coefficient vanishes. The linearity $Q_{\rm nem} \propto \rho_0$ is completely equivalent to the linearity found in the
lattice model in the absence of a substrate potential (see Eq.~(\ref{eq:Q_smallrho_lat})). 

The linearized equation (\ref{eq:feqlin}) is connected to an approximated  free energy per particle $a_{\rm lin}$ 
through $\delta a_{\rm lin}/\delta f_1 =0$.
$a_{\rm lin}$ is quadratic in $f_1$ and is defined to give the
the difference to the isotropic state:
\bea
  a_{\rm lin} &=& \frac{1}{2}  \int_0^{\pi/2} d(\cos \theta) f_1(\theta)^2 + \\ \nonumber
              & &  \frac{\rho_0}{2} \int_0^{\pi/2} d(\cos \theta) \int_0^{\pi/2} d(\cos \theta') \beta_\phi ( \theta, \theta') (2+f_1(\theta)) f_1(\theta') \;.
\eea
For the leading order solution (\ref{eq:lo}), the free energy can be explicitly evaluated.
It is convenient to use the Legendre expansion of the solution: $f_1 = \sum_{i=1}^\infty f_{1,i} P_{2i}(\cos\theta)$
with $f_{1,i} = -(2i+1) \rho_0 B_{i0}$. Using furthermore $B_{i0} = B_{0i}$ one finds
\bea
  a_{\rm lin} & \approx & - \frac{1}{2} \sum_{i=1}^\infty (2i+1) \rho_0^2 B_{i0}^2 \;,
\eea
i.e. the free energy in the anisotropic state is always lower than in the isotropic state.

For hard spherocylinders in the limit $L/D \to \infty$ the excluded area does not depend on whether the rod centers or ends are fixed
to the plane. Through geometric arguments (see App.~\ref{app:excl}) we find
\bea
 \label{eq:exclarea}
  \beta(\theta_1,\theta_2,\phi_2) &=& \frac{2 L D}{\cos\theta_{\rm min}} | \sin \gamma | \;,  \\
   & & \theta_{\rm min} = {\rm min}(|\theta_1|,|\theta_2|) \nonumber \\
   \nonumber            & & \cos\gamma = \cos\theta_1 \cos\theta_2 + \sin\theta_1 \sin\theta_2 \cos\phi_2 \;,
\eea
where $\gamma$ is the angle between the rods. 
Thus we see that for long rods a scaling $Q_{\rm nem} \propto L \rho_0$ is predicted, 
which is different from $Q \propto L^2 \rho$ found in the
lattice model. The numerical evaluation of the Legendre coefficient in Eq.~(\ref{eq:Qnem}) gives 
\bea
  \label{eq:QnemDFT}
  Q_{\rm nem} \approx 0.45\;L D\rho_0 \qquad (L/D \to \infty)\;.
\eea

\subsection{Simulations}

{In order to validate the predictions from the previous section, we have performed Monte Carlo (MC) simulations of hard spherocylinders 
(cylinders of length $L$ capped with two hemispheres of diameter $D$ on either end)
whose centers are restricted to move within a plane (off lattice) while the orientation vectors can take any direction in 
three--dimensional space. A cuboid simulation box with periodic boundary conditions and dimensions $L_{x} \times L_{y} \times (L+D)$  
has been used. Configurations have been generated using single particle displacement and rotations via the Metropolis scheme \cite{Metropolis53} 
as well as a specialized move for small densities
{that forces particles to come close to each other. 
We pick two random particles and move one into a circle of radius $(L+D)$ around the center of mass of the other. In order to impose detailed balance, 
the acceptance rate of the move then simply needs to be multiplied by the ratio between the area of the simulation plane ($L_{x} \times L_{y}$) 
and the area of that circle ($\pi (L+D)^2$).}
We generated configurations for a fixed particle number $N$ while varying the dimensions of the plane $L_{x} \times L_{y}$ to change the area number density of the rods 
{$\rho_0$}. After equilibration, we generated at least $10^{6}$ independent configurations for each value of $L/D$, $N$ and {$\rho_0$} to evaluate the nematic order parameter $Q_{\rm nem}$. }

\begin{figure}
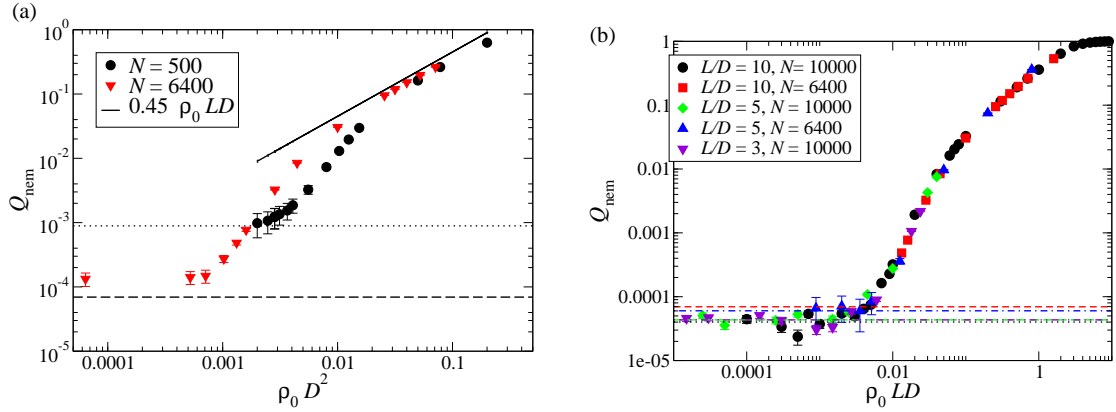

\epsfig{file=fig6a.eps, width=7cm}
   \hspace{5mm}
   \epsfig{file=fig6b.eps, width=7cm}
  \caption{(color online) {(a) Finite system size analysis, rod length $L/D = 10$. The dashed and dotted horizontal lines are $Q_{\rm nem}$ values obtained for a system of freely penetrable rods for particle numbers $N$ = 6400 and 500, respectively, while the solid line is the DFT result (Eq.~(\ref{eq:QnemDFT})). \newline  
   (b) Order parameter $Q_{\rm nem}$ vs. {$\rho_0 (L/D)$} plot for different aspect ratios $L/D$ and numbers of rods $N$. The horizontal lines are as in (a).
    }}
  \label{fig:fixz_wrt_AR&N}
\end{figure}

{At low densities, the simulations are subject to strong finite size effects, which produce artefacts that might be 
misinterpreted as traces of a phase transition. In Fig.~\ref{fig:fixz_wrt_AR&N}(a), we show {$Q_{\rm nem}(\rho_0)$ at low densities for two different $N$.} 
Here, the horizontal lines mark the $Q_{\rm nem}$--values obtained in simulations of freely penetrable rods. For an infinite number of particles, this value would be zero. 
However, {since} in the simulations we sum the orientational order tensor over a finite number of particles its eigenvalues are not exactly zero
{and therefore the largest eigenvalue, corresponding to $Q_{\rm nem}$, is always larger than zero.}
%. When we then order them in order to determine the largest eigenvalue, $Q_{\rm nem}$, obviously we always obtain a value larger than zero.
(This problem is not solved by the common strategy of taking twice the middle eigenvalue instead of the largest eigenvalue. It is inherent {in the restriction to} 
finite particle numbers.) The horizontal lines thus mark the limit of detection of a $Q_{\rm nem}$ that {truly signals orientational anisotropy} for a given number 
of particles $N$. {For a given $N$, there is a first lower density above which orientational order is detectable, 
and a second density where the theoretically expected
behavior $Q_{\rm nem} \propto \rho_0$ sets in. These two densities are particle number dependent and shift towards zero with increasing $N$, and thus are}
not signatures of an additional phase transition.
The solid line in Fig.~\ref{fig:fixz_wrt_AR&N}(a) is the DFT result (Eq.~(\ref{eq:QnemDFT})) derived in the previous section.
{For densities beyond the second density, the simulation results are very close to the DFT result and the density range where this occurs becomes larger with increasing
system size.} 
}

The linearity $Q_{\rm nem} \propto \rho_0$ can also be seen in numerical results for a monolayer of ellipsoids \cite{Var16} (Fig. 4 therein, for an aspect ratio of 10).
The density functional used in Ref.~\cite{Var16} reduces to Eq.~(\ref{eq:fex_hardrods}) in the low--density limit and should therefore comply with the present analysis,
however, explicit expressions have not been given in Ref.~\cite{Var16}. Corresponding Monte--Carlo simulation results in Ref.~\cite{Var16} show agreement
with the DFT results but low densities have not been considered. 

{In Fig.~\ref{fig:fixz_wrt_AR&N}(b), we show $Q_{\rm nem}$ vs. {$\rho_0(L/D)$}, which is {independent of} the aspect ratio $L/D$.}
{System sizes are very similar here such that the finite size effect discussed above is not visible.}
% (and system sizes $N$ for larger values of $\rho$ than discussed above) as predicted.}

\begin{figure}
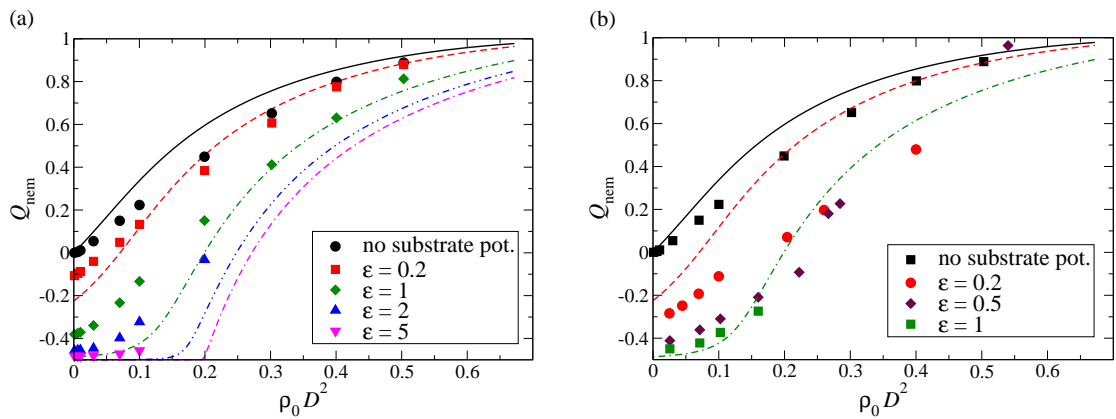

  \epsfig{file=fig7a.eps, width=7cm}  \hspace{5mm}
  \epsfig{file=fig7b.eps, width=7cm}
  \caption{(color online) {Order parameter $Q_{\rm nem}$ vs. number density {$\rho_0 D^2$} for rods subject to an attractive substrate potential ($L/D = 5$). 
The dashed lines are the results from the lattice model shown in Fig.~\ref{fig:2d3d_vext} while data points 
correspond to the off--lattice simulation (errors smaller than symbol size). 
The substrate potential is orientation--dependent defined as (a) $V^{\rm ext} = -\epsilon (L/D)\sin\theta$ and, (b) $V^{\rm ext} = -\epsilon(L/D)\sin^{2}\theta$, 
where $\theta$ is the angle {between rod director and the substrate normal}.}}
  \label{fig:2d3d_off-lattice_Vext}
\end{figure}

{ In the presence of an attractive substrate potential, we {qualitatively} observe the same behaviour as in the lattice model 
(see Fig.~\ref{fig:2d3d_vext}). We compare two different potentials {(a) $V^{\rm ext} = -\epsilon (L/D)\sin\theta$   and, (b) $V^{\rm ext} = -\epsilon(L/D)\sin^{2}\theta$, 
where $\theta$ is the angle between rod director and the substrate normal. 
Both choices drive the system towards nematic order (rods ``standing up"), with choice (b) the external free energy per particle (Eq.~(\ref{eq:aext})) becomes
\bea
  a^{\rm ext} &=& \int_0^{\pi/2} d\theta  \; f(\theta) V^{\rm ext}(\theta) = \frac{2}{3}\,\epsilon\, \frac{L}{D} \, (Q_{\rm nem} -1)\;. 
\eea
Similar to the choice of the external potential in the lattice model (Eq.~(\ref{eq:fext_lat})), the corresponding free energy contribution 
(apart from an additive constant) is proportional to the nematic order parameter.
} 
The plots show qualitative agreement with the lattice model as $Q_{\rm nem}$ remains continuous and the "standing up" transition becomes steeper 
with increasing substrate potential parameter $\epsilon$.}

\section{Summary and discussion}
\label{sec:summary}

In this paper, we have analyzed the equilibrium properties of hard rod monolayers by employing density functional theory and Monte--Carlo simulations
for a cubic lattice model (resulting in restricted translational and orientational degrees of freedom for the rods)  and a continuum model
with hard spherocylinders (having unrestricted in--plane translational and orientational degrees of freedom). We used lattice fundamental measure theory
as a DFT for the lattice model. In two and three dimensional bulk systems, lattice FMT predicts rod demixing and a first oder nematic transition, respectively.
Applied to the monolayer situation, lattice FMT shows a continuous ``standing--up" transition of the rods with increasing density. These results are
in excellent agreement with our results from GCMC simulations. For the continuum model, the same type of continuous ``standing--up" transition
is predicted by DFT in a virial expansion. In MC simulations, the transition is masked by strong finite--size effects but we have evidence that
for large system sizes simulations and DFT agree. Although the transitions in the lattice and the continuum model are very similar, there is a
qualitiative difference in the scaling with the rod extension (at low densities $\rho$). 
The lattice model does not show a simple scaling of the nematic order with respect to  a scaled density variable whereas
in the continuum model the scaling is with $\rho LD$ where $L$ and $D$ are length and diameter of the rods. 
This can be understood from the scaling of the second virial coefficient.     

The presence of an attractive surface potential does not change the continuous character of the ``standing--up" transition. 
However, for
attractive energies per unit rod length which are of the order of 5 $\kt$ or larger, the transition resembles 
a second order transition as present for instance in the bulk 2D system.

Our results are a first step towards modelling the equilibrium and growth of thin films with anisotropic particles with simple coarse grained models.
Investigations of the dynamics of monolayer growth with hard rods would be the logical next step \cite{Klopo16}.
Incorporating particle attractions as well as extending the investigations to multiple layers is desirable and should be pursued both by equilibrium and
growth investigations in order to clarify the influence of the equilibrium phase diagram vs. purely kinetic effects onto the final structures. 

{\bf Acknowledgment:} This work is supported within the DFG/FNR INTER project ``Thin Film Growth" by the Deutsche Forschungsgemeinschaft (DFG), project OE 285/3-1 
and by the Fonds National de la recherche (FNR) Luxembourg. 
Data from computer simulations presented in this paper were produced using the HPC facilities of University
 of Luxembourg \cite{Var14}. The authors acknowledge stimulating discussions with Frank Schreiber.

\begin{appendix}

% \section{Approximative free energy functionals from dimensional crossover}
% \label{app:fmt}

%The central idea is dimensional reduction. Suppose the external potential effectively creates a cavity which can hold only
%one particle

\section{Equilibrium grand--canonical simulations for the monolayer in the lattice model}
\label{app:gcmc}

The simulation results for the lattice monolayer system were obtained using grand canonical Monte Carlo (GCMC) simulations on an $M\times M$ lattice.
We treated the rods with a fixed orientation as a distinct species with corresponding particle number $N_i$ ($i=1...3$). The chemical potential $\mu$ was equal for 
all three species. In each GCMC step, insertion or deletion of a rod was chosen with probability 1/2. Then, the species on which the insertion/deletion is performed, was
chosen with probability 1/3. For the insertion move $N_i \to N_i+1$, a random lattice site was chosen. If no overlap with the existing rods occurs, the move
was accepted with probability $\alpha_{\rm ins} = {\rm min}\left(1, M^2/(N_i+1)\,z\,\exp(-\Delta V^{\rm ext})\right)$ where $z=\exp(\mu/(\kt))$
and $\Delta V^{\rm ext}$ is the change in external energy upon insertion of the rod. 
For the deletion move $N_i+1 \to N_i$, a particle from species $i$ was chosen randomly and removed with probability
$\alpha_{\rm del} = {\rm min}\left(1, (N_i+1)/M^2\,z^{-1}\,\exp(+\Delta V^{\rm ext})\right)$.   

We used lattices with $M=256$ and $10^7$ single moves for a data point with no or small external potential. For stronger external potentials, we used
$10^8$ moves ($\epsilon=\{1,2\}$).

\section{Excluded area for hard spherocylinders in the limit $L/D \to \infty$}
\label{app:excl}

\begin{figure}
 \begin{center}
   \epsfig{file=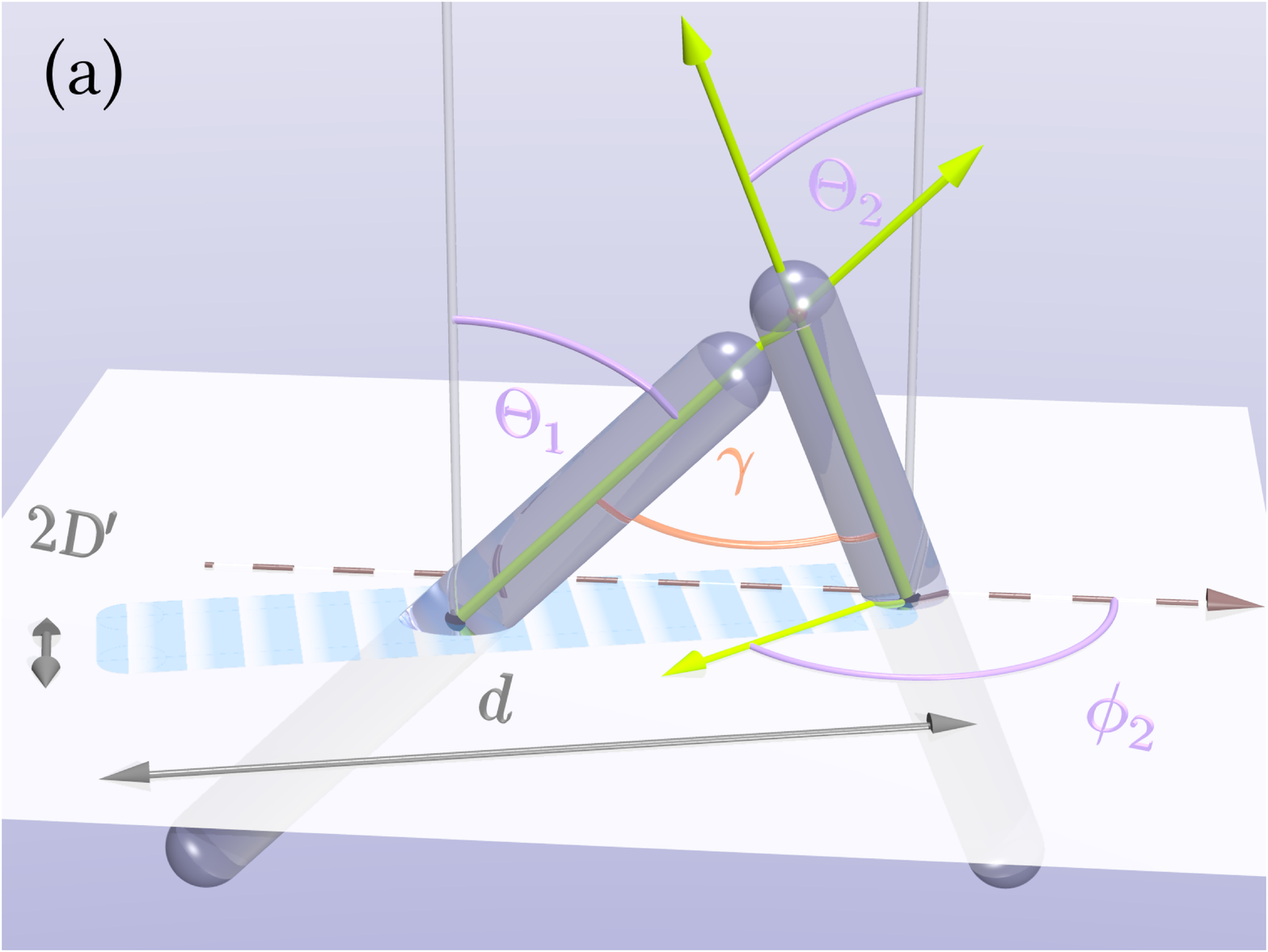,width=4cm} \hspace{5mm} 
   \epsfig{file=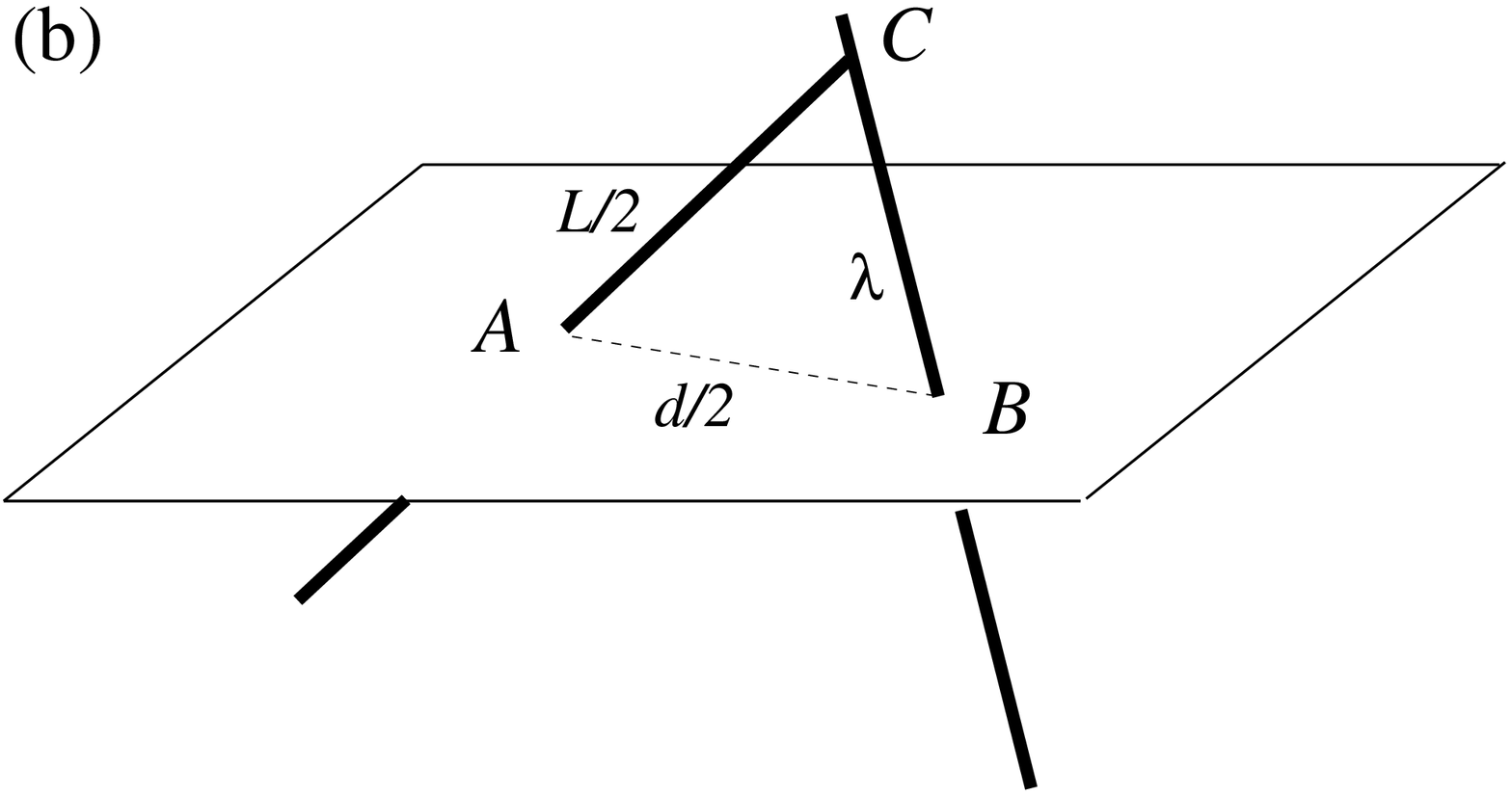,width=4cm} \hspace{5mm}
   \epsfig{file=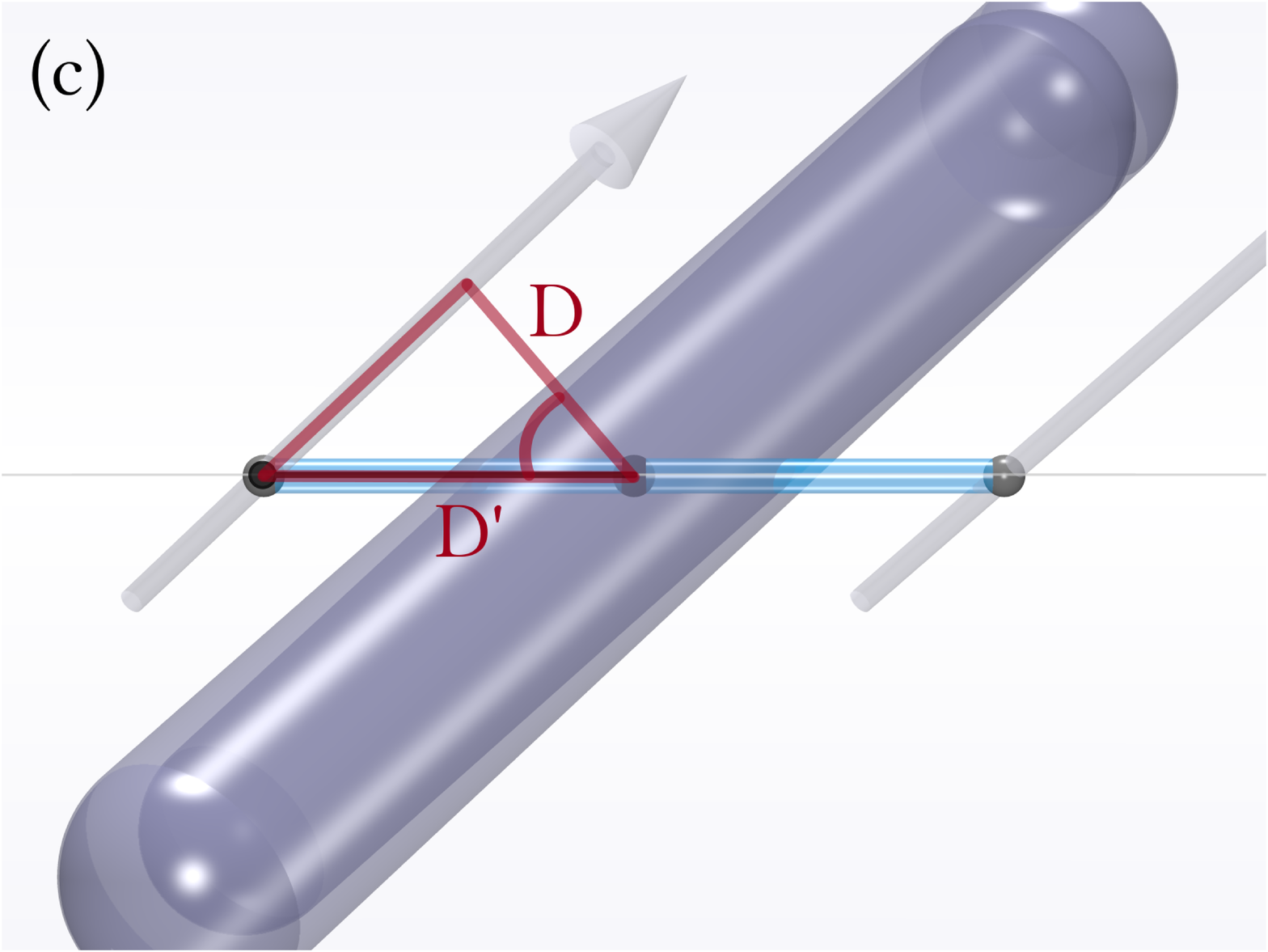,width=4cm}
 \end{center}
 \caption{(color online) (a) and (b): Hard rods (spherocylinders) with their mid points fixed on the substrate plane. The area enclosed in dashed lines is the excluded area and can be approximated
 by a rectangle with side lengths $d$ and $D'$ in the limit $L/D \to \infty$. (c): Side view on the two hard rods from a perspective where rod 1 is exactly
hidden behind rod 2.  
 }
 \label{fig:exclarea}
\end{figure}

Here we briefly derive Eq.~(\ref{eq:exclarea}). The geometric definitions are given in Fig.~\ref{fig:exclarea}. For infinitely thin hard rods,
the (maximum) distance of closest approach $d/2$ is given by
\bea
    d/2 = \frac{(L/2)}{\cos\theta_{\rm min}} \sqrt{\cos^2 \theta_1 + \cos^2 \theta_2 - 2\cos\theta_1 \cos\theta_2 \cos \gamma} \;.
\eea
This is obtained from the law of cosines in the triangle $ABC$ (see Fig.~\ref{fig:exclarea}(b)) where $\lambda=(L/2) \cos\theta_{\rm max}/\cos\theta_{\rm min}$
and $\theta_{\rm min[max]} = {\rm min[max]}(|\theta_1|,|\theta_2|)$. If we consider now a finite, small thickness $D$ of the rods then 
rod 2 may slide past rod 1 at a distance $D'$ to either side of rod 1 along the direction of $d$. This defines the excluded area
(enclosed in dashed lines in Fig.~\ref{fig:exclarea}(a)). It is a rectangle with side lengths $d$ and $2D'$. 
According to Fig.~\ref{fig:exclarea}(c), the distance $D'$ is given by
\bea
 \label{eq:dprime}
   D' = \frac{D}{\sin\alpha} \;,
\eea 
where $\alpha$ is the angle of $\vect e_z$ with the normal vector $\vect n$ to the plane spanned by the two rods. This normal vector is given by
\bea
  \vect n = \frac{ \vect u_1 \times \vect u_2}{|\sin\gamma|}\;,
\eea   
where $\vect u_i$ is the normalized director of rod $i$. Thus
\bea
 \cos\alpha = \vect n \cdot \vect e_z = \frac{\sin\theta_1 \sin\theta_2 \sin\phi_2}{|\sin\gamma|} \;.
\eea
Insertion into Eq.~(\ref{eq:dprime}) and some manipulations give
\bea
   D' = D\; \frac{|\sin\gamma|}{\sqrt{\cos^2 \theta_1 + \cos^2 \theta_2 - 2\cos\theta_1 \cos\theta_2 \cos \gamma}} \;,
\eea
such that finally the excluded area becomes
\bea
  \beta(\theta_1,\theta_2,\phi_2) \approx 2D' d= \frac{2 L D}{\cos\theta_{\rm min}} | \sin \gamma | \;. 
\eea

\end{appendix}

\end{document}